\documentstyle[12pt]{article}

\fnsymbol{footnote}

\newcommand{\be}{\begin{equation}}
\newcommand{\ee}{\end{equation}}
\newcommand{\bea}{\begin{eqnarray}}
\newcommand{\eea}{\end{eqnarray}}

\def\R{{\mathbf R}}

\def\pa{{\partial}}
\def\pa{{\partial}}
\def\tr{{\mathrm tr\,}}       
\begin{document}

\thispagestyle{empty}
\setcounter{page}{0}

\begin{center} 
\Large \bf
On the classical R-matrix of the degenerate Calogero-Moser 
models\footnote{Talk by L. F. at the $8^{th}$ 
Colloquium on Quantum Groups and Integrable Systems, 
Prague, June 1999.} \\
\end{center}

\vspace{.1in}

\begin{center}
L. Feh\'er and B.G. Pusztai\\
\vspace{0.3in}
{\em Department of Theoretical Physics, J\'ozsef Attila University \\
Tisza Lajos krt 84-86, H-6720 Szeged, Hungary \\
e-mail: lfeher@sol.cc.u-szeged.hu }
\end{center}

\vspace{.3in}

\begin{center} {\bf Abstract} \end{center}

The most general momentum independent dynamical r-matrices
are described for the standard Lax representation of the degenerate 
Calogero-Moser models based on $gl_n$ and 
those r-matrices whose dynamical dependence can be 
gauged away are selected. 
In the rational case, a non-dynamical r-matrix 
resulting from gauge transformation is given 
explicitly as 
an antisymmetric solution of the classical Yang-Baxter equation 
that belongs to the Frobenius subalgebra of $gl_n$ consisting of   
the matrices with vanishing last row.

\newpage

\section{Introduction}

The Calogero-Moser type many particle systems \cite{C,M,P}
have recently been the subject of intense studies due to their fascinating 
mathematics and applications ranging from 
solid state physics to Seiberg-Witten theory.
The equations of motion of these classical mechanical systems admit 
Lax representations,
\be
\dot{L}=[L,M], 
\label{1}\ee 
as is necessary for integrability.
More precisely, Liouville ingtegrability also requires \cite{BV} 
that the Poisson brackets of the Lax matrix must be expressable in 
the r-matrix form,
\be
\{ L_1, L_2 \} = \{ L^\mu, L^\nu\} T_\mu \otimes T_\nu = 
[R_{12}, L_1] - [R_{21}, L_2],
\label{2}\ee
where $R_{12}=R^{\mu \nu} T_\mu \otimes T_\nu$ with some 
constant matrices $T_\mu$.

The specification of Calogero-Moser type models typically involves
a root system and a potential function depending on the 
inter-particle `distance'.
The potential is given either by the Weierstrass ${\cal P}$-function or 
one of its (hyperbolic, trigonometric or rational) degenerations.  
A  Lax representation  of the Calogero-Moser  
models based on the root systems of the classical Lie algebras 
was found by Olshanetsky and Perelomov \cite{OP}
using symmetric spaces. 
Recently new Lax representations for these systems 
as well as their exceptional Lie algebraic 
analogues and twisted versions have been constructed \cite{dHP,BCS}.

The classical r-matrix has been explicitly determined in the literature 
in some cases of the Olshanetsky-Perelomov Lax representation 
with degenerate potentials \cite{AT,ABT} and for 
Krichever's \cite{Kri} spectral parameter 
dependent Lax matrix in the standard $gl_n$ case \cite{Skly,BS}. 
The r-matrices turned out to be {\em dynamical}.

Since the present understanding  of most integrable  systems
involves constant r-matrices,
it is natural to ask if the Lax representation 
of the Calogero-Moser models can be chosen in such a way to exhibit
non-dynamical r-matrices.
The obvious way to search for new Lax representations with this property
is to perform gauge transformations on the usual Lax representations
and their dynamical r-matrices.
Our aim here is to implement this program 
in the simplest case: the standard Calogero-Moser models belonging   
to $gl_n$ defined by degenerate potential functions.
A more complete version of our study containing the proofs 
and a comparison with the existing results  on the elliptic case \cite{Hou} 
will be published elsewhere. 
  
\section{The Avan-Talon r-matrix in its general form}

The standard Calogero-Moser-Sutherland models 
are defined by the Hamiltonian
\be
h=\frac{1}{2}\sum_{k=1}^n p_k^2 +  \sum_{k<l} v(q_k-q_l),
\label{3}\ee
where, in the degenerate cases, $v$ is given by 
\be
v(\xi)= \xi^{-2} \quad\hbox{or}\quad
v(\xi)=a^2 \sinh^{-2}(a\xi)
\quad\hbox{or}\quad
v(\xi)=a^2 \sin^{-2}(a\xi).
\label{4}\ee
One has the canonical Poisson brackets
$\{ p_k, q_l\} =\delta_{k,l}$, the coordinates 
are restricted to a domain in $\R^n$ where $v(q_k-q_l)<\infty$,  
and $a$ is a real parameter.

Let us fix the following notation for  elements of the Lie algebra $gl_n$:
\be
H_k:= e_{kk},\,\,
E_{\alpha}:= e_{kl},\,\,
H_\alpha:=(e_{kk}-e_{ll}),\,\,
K_\alpha:= (e_{kk} + e_{ll})
\,\,\hbox{for}\,\,
\alpha =\lambda_k -\lambda_l\in \Phi.
\label{5}\ee
Here $\Phi=\{ (\lambda_k-\lambda_l) \vert k\neq l\}$ is the set of 
roots of $gl_n$, $\lambda_k$ operates on a  diagonal matrix,
 $H={\mathrm{diag}}(H_{1,1},\ldots,H_{n,n})$ as 
$\lambda_k(H)=H_{k,k}$, and $e_{kl}$ is the $n\times n$ elementary 
matrix whose $kl$-entry is $1$. 
Moreover, we denote the standard Cartan subalgebra of 
$sl_n\subset gl_n$ as ${\cal H}_n$,
and put $p=\sum_{k=1}^n p_k H_k$, $q=\sum_{k=1}^n q_k H_k$,
${\bf 1}_n=\sum_{k=1}^n H_k$. 

{}From the list of known Lax representations 
we  consider the original one \cite{C,M} for which $L$ is the $gl_n$ valued 
function  defined by 
\be
L(q,p)=p+ i \sum_{\alpha\in \Phi} w_\alpha(q) E_\alpha,
\label{6}\ee
where $w_\alpha(q):= w(\alpha(q))$ and
$w(\xi)$ is chosen, respectively, as  
\be
w(\xi)= \xi^{-1} \quad\hbox{or}\quad
w(\xi)=a \sinh^{-1}(a\xi)
\quad\hbox{or}\quad
w(\xi)=a \sin^{-1}(a\xi).
\label{7}\ee 
As an $n\times n$ matrix $L_{k,l}= p_k \delta_{k,l} 
+i (1-\delta_{k,l}) w(q_k-q_l)$,
but $L$ can also be used in any other representation of $gl_n$.
The r-matrix corresponding to this $L$ was studied by 
Avan and Talon \cite{AT},
who found that it is necessarily dynamical, and may be chosen 
so as to depend on the coordinates $q_k$ only.
We next describe a slight generalization of their result.

\smallskip 
\noindent {\bf Proposition 1.} {\em 
The most general $gl_n\otimes gl_n$-valued r-matrix 
that satisfies (\ref{2}) with the Lax matrix in (\ref{6}) 
and depends only on $q$ is given by
\be
R(q)= \sum_{\alpha\in \Phi} \frac{w'_\alpha(q)}{w_\alpha(q)}
E_\alpha\otimes E_{-\alpha}
+{1\over 2} \sum_{\alpha\in \Phi}
w_\alpha(q) (C_\alpha(q) - K_{\alpha}) \otimes E_\alpha
+ {\bf 1}_n \otimes Q(q),
\label{8}\ee
where the $C_{\alpha}(q)$ are 
${\cal H}_n$-valued functions subject to the conditions
\be
C_{-\alpha}(q)=- C_\alpha(q),
\qquad
\beta( C_\alpha(q)) = \alpha( C_\beta(q))
\quad
\forall \alpha,\beta\in \Phi 
\label{9}\ee
and $Q(q)$ is an arbitrary $gl_n$-valued function.
}
\smallskip

\noindent
{\em Remarks.}
The functions $C_\alpha$ can be given arbitrarily for the simple roots,
and are then uniquely determined for the other roots by (\ref{9}).
The r-matrix found by Avan and Talon \cite{AT}
is recovered from (\ref{8}) with $C_\alpha\equiv 0$;
and we refer to $R(q)$ in (\ref{8}) as 
the {\em Avan-Talon r-matrix in its general form}. 
If desirable, one may put 
 $Q(q)=\frac{1}{n} \sum_{\alpha\in \Phi} w_\alpha(q) E_\alpha$
to ensure that $R(q)\in sl_n\otimes sl_n$.
Proposition 1 can be proved by a careful calculation
along the lines of \cite{BS}. 

\section{Is $R(q)$ gauge equivalent to a constant r-matrix?}

A gauge transformation of a given Lax representation (\ref{1}) 
has the form 
\be
L \longrightarrow L'=g L g^{-1}
\qquad
M\longrightarrow M' = gM g^{-1} -  {d g\over dt} g^{-1},
\label{10}\ee
where $g$ is an invertible matrix function on the phase space.
If $L$ satisfies (\ref{2}), then $L'$ will have similar Poisson
brackets with a transformed r-matrix $R'$.
The question now is whether one can remove the $q$-dependence of any of the
r-matrices in (\ref{8}) by a gauge transformation of $L$ in (\ref{6}).
It is natural to assume this gauge transformation to be $p$-independent,
i.e.~defined by some function $g:q\mapsto g(q)\in GL_n$, 
in which case 
\be
R'(q)= \left( g(q)\otimes g(q)\right)\left( R(q) + 
\sum_{k=1}^n A_k(q) \otimes H_k\right) 
\left( g(q) \otimes g(q)\right)^{-1}
\label{Rprime}\ee
with 
\be
A_k(q):= -g^{-1}(q) \partial_k g(q), 
\qquad 
\partial_k:= {\partial \over \partial q_k}. 
\label{defA}\ee 
We are looking for a function $g(q)$ for which $\partial_k R'=0$, 
which is equivalent to
\be
\partial_k (R + \sum_{l=1}^n A_l \otimes H_l) 
+[ R + \sum_{l=1}^n A_l \otimes H_l , 
A_k \otimes {\bf 1}_n + {\bf 1}_n \otimes A_k] =0.
\label{13}\ee
By using (\ref{defA}), which means that   
\be
\pa_k A_l - \pa_l A_k + [A_l, A_k]=0,
\label{puregauge}\ee 
it is useful to rewrite (\ref{13}) as 
\be
\pa_k R + \sum_l \pa_l A_k \otimes H_l 
+ [R, A_k \otimes {\bf 1}_n + {\bf 1}_n \otimes A_k ] 
+ \sum_l A_l \otimes [H_l,A_k]=0.
\label{condonA}\ee  
We first wish to solve the last two equations for $A_k$, which 
we now parameterize as
\be
A_k(q)= \sum_{l=1}^n \Psi_k^l(q) H_l + 
\sum_{\alpha\in \Phi} B_k^\alpha(q) E_\alpha.
\label{paramA}\ee
After finding $A_k$ we will have to determine $g(q)$ and 
the resulting constant r-matrix. 

By substituting (\ref{8}) and (\ref{paramA}), from the $E_\alpha\otimes H_k$ 
components of  (\ref{condonA}) we get that
\be
B_k^\alpha(q) = w_\alpha(q) b_k^\alpha, 
\qquad
b_k^\alpha: \hbox{some constants}.
\ee
The $E_\alpha\otimes E_\beta$ components of (\ref{condonA}) 
also do not contain 
the $\Psi_k^l$, and we obtain the following result by detailed inspection.

\smallskip
\noindent
{\bf Proposition 2.}{\em 
The  $E_\alpha\otimes E_\beta$ components of (\ref{condonA}) admit
solution for the constants $b_k^\alpha$ only for those two families 
of $R(q)$ in (\ref{8}) for which the $C_\alpha$ are chosen according to
\be
\hbox{case I}:\quad C_\alpha = - H_\alpha
\quad \forall \alpha\in \Phi, 
\qquad
\hbox{or}\qquad 
\hbox{case II:}\quad 
C_\alpha= H_\alpha 
\quad \forall \alpha\in \Phi.
\label{Ccases}\ee
For $\alpha=\lambda_m-\lambda_l$,  the $b_k^\alpha$ are respectively given by 
\be
 b_k^{(\lambda_m- \lambda_l)}=\delta_{km} + \Omega
\quad 
\hbox{in case I,}\quad\hbox{and}\quad
b_k^{(\lambda_m- \lambda_l)}=\delta_{kl} + \Omega
\quad \hbox{in case II,}
\label{bk}
\label{bcases}\ee
where $\Omega$ is an arbitrary constant.}
\smallskip

Now we present two solutions of the full equations
 (\ref{puregauge}), (\ref{condonA}) making simplifying choices
for the arbitrary function $Q(q)$ in (\ref{8}) and the arbitrary 
constant $\Omega$ in (\ref{bcases}).
 
\smallskip
\noindent
{\bf Proposition 3.}
{\em 
Consider $R(q)$ in (\ref{8}) with $Q=0$ and the $C_\alpha$ in (\ref{Ccases}).
Then a solution of eqs.~(\ref{puregauge}), (\ref{condonA}) 
for $A_k$ in (\ref{paramA}) is provided by $b_k^\alpha$ 
in (\ref{bcases}) with $\Omega=0$, and 
$\Psi_k^l$ defined by  
\be
\Psi_k^k=0, 
\qquad  
\Psi_k^l(q)= -
\frac{ w'(q_l - q_k)}{w(q_l - q_k)}
\quad\hbox{for}\quad
 k\neq l\quad  
\hbox{in case I},
\label{PsiI}\ee
\be
\Psi_k^k=0, 
\qquad 
\Psi_k^l(q)= \frac{ w'(q_l - q_k)}{w(q_l - q_k)}
\quad \hbox{for}\quad
k\neq l
\quad \hbox{in case II.}
\label{bk2}
\label{PsiII}\ee   
}
\smallskip

\noindent
{\em Remark.}
The symmetric part of $R'(q)$
is easily checked to {\em vanish} for either of the two gauge
transformations determined by Proposition 3.
Recall that for an antisymmetric constant $R'$ 
the classical Yang-Baxter equation and 
its modified version are sufficient conditions for the Jacobi identity 
$\{\{ L'_1, L'_2\}, L'_3\} + {\mathrm{cycl.}}=0$.

\section{The constant r-matrix in the rational case}

In general, if $A_k$ is given so that (\ref{puregauge}) holds 
then the gauge transformation $g(q)$ can be determined from 
the differential equation in (\ref{defA}) up to an arbitrary
constant.
We here describe the result in the rational case.
The gauge transformed r-matrix 
found below  is an antisymmetric, constant 
solution of the classical Yang-Baxter equation,
\be
[R'_{12}, R'_{13}] + [R'_{12}, R'_{23}] + [R'_{13}, R'_{23}]=0.
\label{CYB}\ee

Let us consider case I of the preceding  propositions.
Put $w_{kl}$ for $w_\alpha$ with $\alpha=(\lambda_k -\lambda_l)$,
and introduce the notation
\be
I^n_k:= \{ 1,\ldots,n\} \setminus \{ k\}
\qquad 
\forall k=1,\ldots,n.
\ee
Then  $R(q)$ and $A_k(q)$ are the $n\times n$ matrices: 
\be
R=\sum_{1\leq k\neq l\leq n}\left( \frac{w'_{kl}}{w_{kl}}
e_{kl}\otimes e_{lk}
-w_{kl} e_{kk}\otimes e_{kl}\right),
\quad
A_k=\sum_{l\in I^n_k}\left( w_{kl} e_{kl} - 
\frac{w'_{lk}}{w_{lk}} e_{ll}\right).    
\label{RA}\ee
Define now an $n\times n$ matrix $\varphi(q)$ as follows:
$\varphi_{nk}(q):=1$ for any $k=1,\ldots, n$ and 
\be
\varphi_{jk}(q):=\sum_{P (P\subset I^n_k, \vert P\vert=n-j) }
\left( \prod_{l\in P} q_l \right) 
\qquad\qquad
\forall k, \quad 
1\leq j \leq n-1. 
\ee
Here $\vert P\vert$ denotes the number of the elements in the
summation subset $P\subset I^n_k$. 
Note that $\varphi(q)$ is invertible, since
\be
\det [ \varphi(q)]=\prod_{1\leq j<k\leq n} (q_k - q_j) \neq 0.
\ee
Our result now is  

\smallskip
\noindent
{\bf Theorem 4.}
{\em A gauge transformation $g(q)$ for which $\pa_k g(q)=- g(q)A_k(q) $ 
with $A_k$ in (\ref{RA}) is given in the rational case by 
$g(q)=\varphi(q)$, where $\varphi(q)$ is defined above.
The corresponding gauge transform of $R(q)$ in (\ref{RA}) is the following
antisymmetric  solution of (\ref{CYB}):
\be
R'= \sum_{(a,b,c,d)\in S}\,
 (e_{ab}\otimes e_{cd} - e_{cd}\otimes e_{ab}),
\label{constR}\ee
$$
S=\{ (a,b,c,d) \vert a,b,c,d \in {\bf Z},\,
a+c+1=b+d,\, 1\leq b\leq a <n,\,
b\leq c <n,\, 1\leq d\leq n \}.
$$
}
\smallskip

The above constant r-matrix is actually very well-known.
It already appears as an example at the end of \cite{BD},
where it has also been identified in terms of a 
non-degenerate $2$-coboundary of a Frobenius 
subalgebra of $gl_n$.  We briefly recall this interpretation next.

Let us define the subalgebra ${\cal F}_n\subset gl_n$ as
\be
{\cal F}_n ={\mathrm{span}}
\{ T_a\,\vert\, T_a=e_{kl} \quad\hbox{for}\quad  
1\leq k\leq n-1,\quad 1\leq l\leq n\,\}.
\ee
That is, ${\cal F}_n$ consists of the $n\times n$ matrices having zeros 
in their last row.
It is clear that $R' \in {\cal F}_n \wedge {\cal F}_n$, i.e., 
with the basis $T_a$ of ${\cal F}_n$ one can write
\be
R'= \sum_{a,b} {\cal M}_{a,b} (T_a \wedge T_b).
\ee
It is then easy to verify that the matrix ${\cal M}_{a,b}$,
whose size is ${\mathrm{dim}}({\cal F}_n)=n(n-1)$, is invertible, 
and its inverse 
is given by 
\be
{\cal M}^{-1}_{a,b}= \Lambda_n([ T_a, T_b]),
\ee
where $\Lambda_n$ is the linear functional on ${\cal F}_n$ 
defined by 
\be
\Lambda_n(T):= \tr(J_n T)
\quad \forall T\in {\cal F}_n\quad 
\hbox{with}\quad
J_{n}:= \sum_{k=1}^{n-1} e_{k+1,k}.
\label{PLambda}\ee
This realization of $R'$ means that it indeed   
belongs to the Frobenius subalgebra ${\cal F}_n\subset gl_n$, 
and the corresponding inverse is the non-degenerate 2-coboundary 
obtained from  
the functional $\Lambda_n$ in (\ref{PLambda}).

It is interesting to notice 
that $J_n$ in (\ref{PLambda}) is a principal nilpotent element of $gl_n$.
This fact could perhaps be related to a possible interpretation 
of $R'$ as a `boundary solution' of (\ref{CYB}) in the sense of \cite{GG},
which may in turn be related to the degeneration of 
the hyperbolic/trigonometric
Calogero-Moser models into the rational ones.
In the future,
we hope to report on this question as well as on the possible relationship 
of $R'$ in (\ref{constR}) to Belavin's  elliptic r-matrix, which occurs 
for the elliptic Calogero-Moser models according to \cite{Hou}.
The results in \cite{H,AHZ,ACF} concerning 
non-dynamical R-matrices for quantized Ruijsenaars models 
may also be relevant to answer these questions.

\bigskip
\noindent
{\small 
{\bf  Acknowledgments.}
L.F. wishes to thank A. Stolin for a useful discussion.
This work has been supported in part by the Hungarian Ministry of Education
under  FKFP 0596/1999 and by the National Science Fund (OTKA) under T025120. 
}

\end{document}